\documentclass[twocolumn,aps,pra,groupedaddress]{revtex4}

\usepackage{graphicx}
\usepackage{dcolumn}
\usepackage{bm}
\usepackage{amssymb}
\usepackage{amsfonts}
\usepackage{subfig}
\usepackage{latexsym}
\usepackage{enumerate}
\usepackage[dvipdfmx]{color} 
\usepackage{amsmath}
\usepackage[dvipdfmx]{epsfig}
\usepackage{times}
\usepackage{caption}

\newtheorem{theorem}{Theorem}
\newtheorem{lemma}{Lemma}
\newtheorem{remark}{Remark}

\begin{document}
\widetext
\title{Fault-tolerant verifiable blind quantum computing
with logical state remote preparation}
\author{Yuki Takeuchi,${}^{1,\ast}$ Keisuke Fujii,${}^{2,3}$ Tomoyuki Morimae,${}^{3,4}$ and Nobuyuki Imoto${}^1$}
\affiliation{$^1$Graduate School of Engineering Science, Osaka University, Toyonaka, Osaka 560-8531, Japan\\
$^2$Photon Science Center, Graduate School of Engineering, The University of Tokyo, 2-11-16 Yayoi, Bunkyo-ku, Tokyo 113-8656, Japan\\
$^3$JST, PRESTO, 4-1-8 Honcho, Kawaguchi, Saitama 332-0012, Japan\\
$^4$Department of Computer Science, Gunma University, 1-5-1 Tenjin-cho, Kiryu, Gunma 376-0052, Japan}
\affiliation{$$}
\affiliation{${}^\ast${\rm takeuchi@qi.mp.es.osaka-u.ac.jp}}

\begin{abstract}
Verifiable blind quantum computing allows a client 
with poor quantum devices 
to delegate universal quantum computing to a remote
quantum server in such a way that the client's privacy is protected and
the honesty of the server is verified.
In existing protocols, the client has to send single-qubit states 
to the server. 
These states might be decohered by the channel noise.
Furthermore, the client hides some ``trap" qubits in the server's register
so that the client can detect the server's deviation.
In reality, however, these trap qubits are disturbed by imperfect operations by the server, which reduces the probability that the client accepts the honest server.
To solve these problems, 
we propose a new gadget that allows the client to remotely
prepare encoded logical single-qubit states in the server's place.
Importantly, in our fault-tolerant verifiable blind quantum computing protocol, the client needs only the ability of physical single-qubit
measurements in $X$ and $Z$ bases.
\end{abstract}

\maketitle 

\section{INTRODUCTION}
Because of its high maintenance,
a first-generation quantum computer would be realized 
in a ``cloud style": 
a client with poor quantum devices delegates 
universal quantum computing to a remote quantum server.
In such a cloud quantum computing,
protecting client's privacy is 
of prime importance.
Blind quantum computing (BQC) protocols guarantee blindness, i.e.,
information-theoretic
security of the client's input, quantum algorithm, and output.
So far, various BQC protocols~\cite{[ABE10],[BFK09],[FK12],[DKL12],[MF12],[M12],[MF13A],[SKM13],[MF13L],[GMMR13],[MDF13],[BFKW13],[KKD14],[LCWW14],[M14],[K14],[SZ15],[HDF15],[MDK15],[TFIYI15],[DF15],[GKW15],[XL15],[KW15],[HM15],[HH16],[DK16],[FH17],[ABEM17],[FKD17]} have been proposed.
In particular, the Broadbent-Fitzsimons-Kashefi (BFK) protocol~\cite{[BFK09]}, which is based on measurement-based quantum computation (MBQC)~\cite{[RB01]}, has successfully allowed the client to be almost classical. Subsequently, the client's quantum ability~\cite{[BFK09],[DKL12],[MF13A],[MF13L],[LCWW14],[SZ15],[MK14],[DK16]}, communication complexity~\cite{[GMMR13],[MDF13],[DF15]}, composable security~\cite{[MK13],[DFPR14]}, and applications~\cite{[SYWX15],[LCZ16]} of BQC have been studied. Proof-of-principle experiments for several BQC protocols have already been demonstrated using four photonic qubits~\cite{[BFKW13],[BKBFZW12],[GRBMW16]}.

In addition to the information-theoretic security, 
there is another important requirement,
namely, the verifiability, which means that the client can 
verify whether the server honestly performed the delegated quantum computing or not.
In fact, verification methods~\cite{[FK12],[BFKW13],[KKD14],[M14],[HDF15],[GKW15],[KW15],[HM15],[HH16],[FH17],[FKD17]} of BQC have been actively studied.
They are important not only in the cryptographic context,
but also for the understanding of the foundation of 
quantum physics~\cite{[ABE10],[ABEM17]}.
Experimentally verifying the correctness of a physical theory
is essential in physics, but
verifying a quantum many-body theory is a non-trivial task
due to the high complexity of quantum many-body systems.
Verification methods of BQC are nice theoretical
models for studying such a problem.

BQC combined with the verification protocol is called verifiable BQC (VBQC). Fitzsimons and Kashefi have proposed a VBQC protocol, which is called 
the FK protocol~\cite{[FK12]}. In the FK protocol, the client generates ten kinds of single-qubit states 
\begin{eqnarray*}
\{|0\rangle,|1\rangle\}\cup
\{(|0\rangle+e^{ik\pi/4}|1\rangle)/\sqrt{2}~|~0\le k\le 7,k\in\mathbb{Z}\}
\end{eqnarray*}
and sends them to the server. The sever entangles them with the controlled-$Z$
$(CZ)$ gates to prepare an appropriate graph state, which is used for MBQC. Since some of states generated by the client are the $Z$-basis states and are not entangled by the server's $CZ$ gates, some single-qubit states surrounded by the $Z$-basis states are isolated from the graph state, which are called trap qubits. Accordingly, the client can completely predict measurement outcomes on trap qubits.
On the other hand, the server does not know which qubits are trap qubits. As a result, if the server attempts to perform deviation, the server ends up disturbing the state of trap qubits with high probability. Therefore, the client can verify whether the server follows the correct procedure or not by checking outcomes of single-qubit measurements on trap qubits (See Appendix A for the detail of the FK protocol).

One problem of
the existing VBQC protocols based on the trap technique~\cite{[FK12],[BFKW13],[KKD14],[HDF15],[GKW15],[KW15]}
is that they are not fault-tolerant.
If the client sends the bare single-qubit states to the server, they decohere
in the quantum channel from the client to the server.
Another problem of using bare qubits is that 
if trap qubits are not logically encoded, even
the honest server is rejected by the client since
in reality the server's operations are imperfect.
If the client could generate 
and send
{\it logically encoded} ten kinds of states,
\begin{eqnarray}
\{|0_L\rangle,|1_L\rangle\}\cup
\{(|0_L\rangle+e^{ik\pi/4}|1_L\rangle)
/\sqrt{2}~|~0\le k\le 7,k\in\mathbb{Z}\},
\label{ten}
\end{eqnarray}
to the server, the fault-tolerance is maintained, but
it is unrealistic since the client has to perform
entangling operations.

In this paper, to solve the problem, we propose a new
gadget that allows the client to remotely prepare
the ten logical states of Eq.~(\ref{ten}) in the server's place
in such a way that the server cannot learn which states
are prepared.
These logical single-qubit states are encoded in the
Calderbank-Shor-Steane (CSS) code~\cite{[CS96],[S96]}.
Importantly, in the gadget, the client needs only
the ability of physical single-qubit measurements in $X$ and $Z$ bases.
We construct a fault-tolerant VBQC protocol by combining the gadget to the FK protocol.
Since the client of the FK protocol needs no quantum operation after sending ten kinds of states to the server, thus constructed fault-tolerant VBQC protocol requires the client to have only the ability of single-qubit measurements in $X$ and $Z$ bases.

\begin{figure}[t]
\begin{center}
\includegraphics[width=6cm, clip]{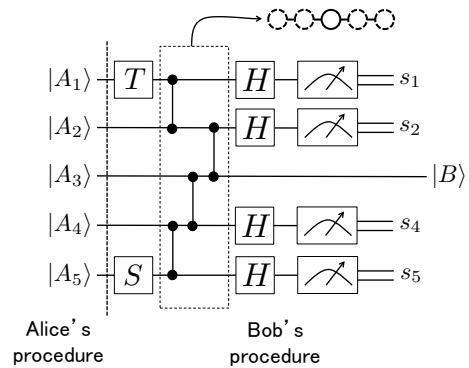}
\end{center}
\captionsetup{justification=raggedright,singlelinecheck=false}
\caption{The quantum circuit used in our gadget. For simplicity, we omit the subscript $L$. In the graph state representation at the top, solid and dashed circles indicate the output state $|B\rangle$ and the measured qubits, respectively.}
\label{mds_graph}
\end{figure}

An intuitive idea of our gadget is as follows
(For details, see Sec.~\ref{II}).
First, if the server is honest, he 
generates logical Bell pairs and sends one half of each of them
to the client.
Thanks to the transversality of the CSS code, 
the client can prepare logical $X$- and $Z$-basis states (up to correctable errors) in the server's 
place by only physical single-qubit measurements in $X$ and $Z$ bases.
Since non-Clifford measurements cannot be done in the transversal way, 
all of logical states in Eq.~(\ref{ten}) cannot be prepared in 
the server's place in this way.
We therefore introduce our new protocol that enables
the server to generate ten logical single-qubit states of Eq.~(\ref{ten})
from logical $X$- and $Z$-basis states (Details of this protocol is explained in Sec.~\ref{II}, and see Fig.~\ref{mds_graph}).
In this way, the client can remotely prepare ten logical
single-qubit states of Eq.~(\ref{ten}) in the client's place
on which they can run the FK protocol.
One might think that the halves of logical Bell pairs could decohere
during the channel from the server to the client.
However, by virtue of the CSS code, the client can correct errors 
via classical processing after the transversal $\{X,Z\}$-basis measurements,
similarly to the Bennett-Brassard (BB84) protocol~\cite{[BB84],[SP00]} for quantum key distribution (QKD).
For example,
if independent $X$ and $Z$ errors occur in the channel, the client's measurement apparatus, and the server's devices, 
the proposed protocol tolerates
an error rate up to $\sim11\%$~\cite{[CS96],[DKP02]} in total.
In other words, 
the acceptance rate 
can be successfully amplified 
by the almost classical client 
even if there are 
the channel noise, and imperfections of the client's measurement apparatus and the server's devices.

By combining our gadget to the FK protocol, we construct
a fault-tolerant VBQC protocol in Sec.~\ref{III}.
Our fault-tolerant VBQC protocol requires the client to 
have the ability of only single-qubit measurements in $X$ and $Z$ bases.
Such a requirement is the minimum one.
One might point out that the client in other BQC protocols 
that use multiple servers~\cite{[BFK09],[MF13L],[SZ15]}
is more classical than ours. 
However, in these protocols,
a massage sent from the client to a 
server should not be leaked to another server. 
To guarantee such a security, information-theoretically secure 
classical communication should be established between the client and 
each server. In order to achieve such a secure classical communication, 
quantum key distribution (QKD) should be ultimately
employed. For example, if BB84~\cite{[BB84]} is used,
the client anyway has to perform $X$- and $Z$-basis measurements. 

Note that in Ref.~\cite{[DK16]}, 
a protocol was proposed that enables the server to generate
eight kinds of single-qubit states 
\begin{eqnarray*}
\{(|0\rangle+e^{ik\pi/4}|1\rangle)/\sqrt{2}~|~0\le k\le 7,k\in\mathbb{Z}\}
\end{eqnarray*}
from two kinds of single-qubit states sent from the client.
The protocol is useful for the BFK protocol, but
not for the FK protocol, since 
the FK protocol needs the $Z$-basis state preparation
in addition to the above eight states. On the other hand, after the first version of this paper appeared on arXiv, a VBQC protocol that utilizes only the above eight states has been proposed~\cite{[FKD17]}. It might be possible to construct a fault-tolerant VBQC protocol by combining results in Refs.~\cite{[DK16],[FKD17]}. It is an interesting open problem.

Our fault-tolerant VBQC protocol is based on the FK protocol. 
There is another type of VBQC protocols that are based on
the stabilizer testing~\cite{[MF13A],[HM15],[HH16],[FH17]}. 
In these protocols, the server sends the client many copies of graph states. The client randomly samples some of copies and check their stabilizers. If all stabilizer measurements give correct values, remaining graph states are guaranteed to be close to the correct graph states.
Recently, Fujii and Hayashi have proposed its fault-tolerant version~\cite{[FH17]}. 
Our protocol does 
not supersede theirs, 
and vice versa,
since the VBQC protocols based on the FK protocol 
and those based on the stabilizer testing are different:
the latter achieves simpler proofs of the verifiability and the stronger security based on the
no-signaling principle, while the former is free from
on-line quantum communication,
i.e., no quantum communication is necessary
after the client decides her algorithm.

The rest of the paper is organized as follows. 
In Sec.~\ref{II}, we explain our new gadget to remotely prepare 
ten logical states in the server's place. 
In Sec.~\ref{III}, as a main result of the present paper, 
we construct a fault-tolerant VBQC protocol by combining our gadget 
and the FK protocol. 
In the same section, we show its fault-tolerance and discuss its loss-tolerance. 
We then show the correctness (Sec.~\ref{IV}), blindness (Sec.~\ref{V}), 
and verifiability (Sec.~\ref{VI}) of our VBQC protocol.

\medskip
\section{Gadget}
\label{II}
In this section, we explain our gadget to remotely prepare ten logical states in Eq.~(\ref{ten}).
Our gadget runs as follows:
\begin{enumerate}
\item Alice (the client) randomly chooses five bits $(c_1,c_2,c_3,c_4,c_5)\in\{0,1\}^{\times 5}$, where $c_i$ is chosen to be $0$ with probability $q_i$ for each $i=1,2,\dots,5$. Here, $q_1=q_5=p/(1-p)$ $(0<p<1/2)$, $q_2=q_4=1-p$, and $q_3=1-p'$ $(0<p'<1)$. Note that $p$ and $p'$ are specified later. Next, she chooses two sets of five bits $(a_1,a_2,a_3,a_4,a_5)\in\{0,1\}^{\times5}$ and $(r_1,r_2,r_3,r_4,r_5)\in\{0,1\}^{\times5}$ independently and uniformly random.  

\item Alice and Bob (the server) repeat the following steps for $i=1,2,\dots,5$.
\begin{enumerate}
\item[2-a.] Bob sends Alice one half of the logical Bell pair
\begin{eqnarray*}
|\Phi_L^+\rangle\equiv\cfrac{|0_L0_L\rangle+|1_L1_L\rangle}{\sqrt{2}}
\end{eqnarray*}
encoded in the CSS code with length $l$ through a quantum channel.

\item[2-b.] If $c_i=0$, Alice measures the $i$th logical qubit sent from Bob in the $Z_L=Z^{\otimes l}$ basis.  After that, she performs error correction through classical processing to obtain the $i$th reliable measurement outcome $o_i$. She then requests Bob to perform $X_L^{a_i\oplus o_i}Z_L^{r_i}$ on his $i$th half.

On the other hand, if $c_i=1$, she measures in $X_L=X^{\otimes l}$ and requests Bob to perform $X_L^{r_i}Z_L^{a_i\oplus o_i}$ on his $i$th half.

Now Bob has
\begin{eqnarray*}
|A_{i,L}\rangle\equiv H_L^{c_i}X_L^{a_i}|0_L\rangle,
\end{eqnarray*}
where $H_L=H^{\otimes l}$ is the logical Hadamard gate.
\end{enumerate}

\item 
Bob implements a quantum circuit composed of $S_L\equiv\sqrt{Z_L}$, $T_L\equiv\sqrt{S_L}$, $H_L$, 
$\Lambda(Z_L)$, and $Z_L$-basis measurements, as shown in Fig.~\ref{mds_graph}. Here, $\Lambda(Z_L)$ is the logical $CZ$ gate. 
He then obtains measurement outcomes $(s_1,s_2,s_4,s_5)\in\{0,1\}^{\times 4}$. Let the state of the 3rd output qubit of the circuit in Fig.~\ref{mds_graph} be $|B_L\rangle$. The explicit form of $|B_L\rangle$ depends on $\{a_i\}$, $\{c_i\}$, and $\{s_i\}$ (See Table~\ref{B_i}). 
He sends $s_1$, $s_2$, $s_4$, and $s_5$ to Alice through a classical channel. 
If $s_1=s_2=s_4=s_5=0$, he keeps $|B_L\rangle$. Otherwise, he discards it.
\end{enumerate}
In Table~\ref{B_i} and hereafter, we define 
\begin{eqnarray*}
|+_{k,L}\rangle\equiv(|0_L\rangle+e^{ik\pi/4}|1_L\rangle)/\sqrt{2}\ \ \ (0\le k\le 7, k\in\mathbb{Z}).
\end{eqnarray*}

\begin{center}
\begin{table}[t]
\begin{tabular}{|c|c|c|} \hline
& $(c_1,c_2,c_3,c_4,c_5)$ & $|B_L\rangle$ \\ \hline\hline
(1) & $(0/1,0/1,0,0/1,0/1)$ & $X_L^{a_3}|0_L\rangle$ \\ \hline
(2) & $(0,1,1,0/1,0/1)$ & $X_L^{a_1\oplus a_2}|0_L\rangle$ \\ \hline
(3) & $(0/1,0,1,1,0)$ & $X_L^{a_4\oplus a_5}|0_L\rangle$ \\ \hline
(4) & $(1,1,1,1,0)$ & $X_L^{a_4\oplus a_5}|0_L\rangle$ \\ \hline
(5) & $(0/1,0,1,0,0/1)$ & $Z_L^{a_2\oplus a_3\oplus a_4}|+_{0,L}\rangle$ \\ \hline
(6) & $(0/1,0,1,1,1)$ & $Z_L^{a_2\oplus a_3\oplus a_4\oplus a_5}|+_{2,L}\rangle$ \\ \hline
(7) & $(1,1,1,0,0/1)$ & $X_L^{a_2}Z_L^{a_1\oplus a_3\oplus a_4}|+_{1,L}\rangle$ \\ \hline
(8) & $(1,1,1,1,1)$ & $X_L^{a_2}Z_L^{a_1\oplus a_2\oplus a_3\oplus a_4\oplus a_5}|+_{3,L}\rangle$ \\ \hline
\end{tabular}
\captionsetup{justification=raggedright,singlelinecheck=false}
\caption{The explicit form of $|B_L\rangle$ when $s_1=s_2=s_4=s_5=0$. Here, $0/1$ means that $0$ or $1$.}
\label{B_i}
\end{table}
\end{center}
  
\medskip
\section{Fault-tolerant VBQC protocol}
\label{III}
In this section, as the main result of this paper, we propose a fault-tolerant VBQC protocol by incorporating our gadget in the FK protocol (See also Fig.~\ref{ours}). It runs as follows:
\begin{enumerate}
\item Let $N_D$ be the number of logical $Z$-basis states used in the FK protocol. Let $(N-N_D)$ be that of states $\{|+_{k,L}\rangle\}$ used in the FK protocol. Alice and Bob run the gadget given in Sec.~\ref{II} with $p$ and $p'$ chosen such that
\begin{eqnarray*}
\cfrac{N_D}{N}=1-4p^2(1-p').
\end{eqnarray*}

\item Alice and Bob repeat step $1$ until $N_D$ logical $Z$-basis states and $(N-N_D)$ states $\{|+_{k,L}\rangle\}$ are prepared at Bob's side~\cite{[FN1]}.

\item Alice and Bob perform the FK protocol using logical qubits prepared in step $2$ (See Appendix A for the detail of the FK protocol).

\begin{figure}[t]
\begin{center}
\includegraphics[width=7.5cm, clip]{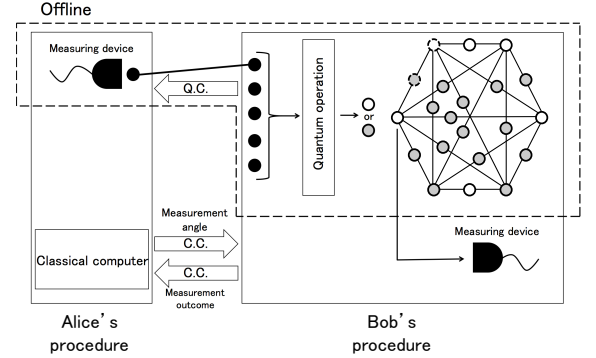}
\end{center}
\captionsetup{justification=raggedright,singlelinecheck=false}
\caption{Our fault-tolerant VBQC protocol. Here, the quantum operation represents the quantum circuit shown in Fig.~\ref{mds_graph}. A white colored and a gray colored circles represent a state $|+_{k,L}\rangle$ and a logical $Z$-basis state, respectively. Two black colored circles connected with each other represent a logical Bell pair. Q.C. and C.C. are abbreviations of quantum communication and classical communication, respectively.}
\label{ours}
\end{figure}

\end{enumerate}

As mentioned earlier, VBQC protocols have to satisfy the blindness and verifiability. In addition to them, VBQC protocols should also satisfy the correctness, which means that if the client and the server follow the correct procedure, the client can obtain the correct output. Our fault-tolerant VBQC protocol indeed satisfies these three requirements, as we show later in Secs.~\ref{IV}, \ref{V}, and \ref{VI}.

Now we explain the fault tolerance of our protocol. By virtue of the error correction,
we can amplify the acceptance rate 
even in the presence of Bob's imperfection and the quantum channel noise.
Note that when we argue the fault tolerance, Bob is assumed to be honest and follows the correct procedure, because otherwise Bob can perform any deviation and therefore the fault tolerance is trivially impossible.

Let us consider the simplest case where errors occur independently in the quantum channel and the server's devices with probability $p_{\rm error}$.
Without error correction, 
the acceptance rate decreases as
$O((1-p_{\rm error})^{N/3} )$ in the FK protocol~\cite{[FK12]}.
On the other hand, in our fault-tolerant VBQC protocol, qubits are always encoded into an error-correcting code.
Let $p_L<e^{-\kappa}$ be the logical error probability per elementary operation, whose overhead is at most a polynomial function of $\kappa$.
Since the number of operations is at most ${\rm poly}(N)$, the acceptance rate under error correction 
becomes 
$O((1-{\rm poly}(N)p_L)^{N/3})\sim O(e^{{\rm poly}(N)p_L {N/3}})$ (More rigorously, according to fault-tolerant theory, we can simulate ideal quantum computing with an exponentially small additive error with respect to $l_1$ norm with a polynomial overhead if the amount of noise measured, for example, by the diamond norm is sufficiently smaller than a certain threshold value).
That is, if we want to satisfy $p_L<O(1/{\rm poly}(N))$, we can amplify the acceptance rate
using a polylog overhead with respect to $N$
as long as Bob's imperfection and the quantum channel noise are small enough.
For clarity, let us consider the case, where 
$X$ and $Z$ errors are
introduced independently with probability $p_{\rm error}$ as channel noise.
If $p_{\rm error} < 11\%$~\cite{[CS96],[DKP02]},
$p_L$
can be reduced exponentially with $\kappa$.
Not only the channel noise, but also errors at Bob's operation can also be made fully fault-tolerant by doing the FK protocol using logical qubits in a fault-tolerant way~\cite{[FY10],[FY10R]}.
While we here consider a specific error model, a similar argument holds in general.
If Bob's deviation or errors are correctable, the acceptance rate is amplified close to unit. Otherwise, the verification protocol automatically rejects Bob's output.

Furthermore, we consider an effect of loss in the quantum channel. Since a logical qubit sent from Bob to Alice is composed of ${\rm polylog}(N)$ qubits, our fault-tolerant VBQC protocol is not efficient for a lossy quantum channel. To make it efficient for loss, we modify our gadget as follows: First, if Alice wants to prepare a logical $X(Z)$-basis state at Bob's side, she measures one half of a bare Bell pair $|\Phi^+\rangle$ sent from Bob in the $X(Z)$-basis until $l$ qubits are prepared at Bob's side. Then, she tells Bob which qubits are reached at her side. Second, Bob generates $|\Phi^+_L\rangle$ at his side. Then, Bob performs quantum teleportation on one qubit of logical one half of $|\Phi^+_L\rangle$ and a remaining one half of $|\Phi^+\rangle$, whose another one half reaches Alice's side, $l$ times. Finally, according to measurement outcomes of Alice's measurements and Bob's quantum teleportations, she requests Bob to perform the logical Pauli operator as with the original gadget. As a result, one logical qubit is prepared at Bob's side as with the original gadget. This modification decreases the mean number of qubits required to prepare one logical qubit at Bob's side from $(1/p_{\rm loss})^l$ to $l/p_{\rm loss}$. Here, $(1-p_{\rm loss})$ is the transmittance of the quantum channel. Note that hereafter, we assume a loss-less quantum channel for simplicity.

\medskip
\section{Correctness}
\label{IV}
In this section, we show that our fault-tolerant VBQC protocol satisfies correctness. To this end, it is sufficient to show that when Alice and Bob follow the correct procedure, Bob obtains ten kinds of single-qubit states in Eq.~(\ref{ten}). Note that in this section, for the notational simplicity, we omit the subscript $L$ of $|B_L\rangle$, $|+_{k,L}\rangle$, $|0_L\rangle$, and $|1_L\rangle$.
\begin{theorem}
\label{theorem1_2}
If Alice and Bob follow the correct procedure in Sec.~\ref{II}, $\{|+_k\rangle\}_{k=0}^7$, $|0\rangle$, and $|1\rangle$ are each prepared at Bob's side with probability $(N-N_D)/(128N)$, $N_D/(32N)$, and $N_D/(32N)$, respectively.
\end{theorem}
{\it{Proof.}}
First, if Alice and Bob follow the correct procedure in Sec.~\ref{II}, then Bob obtains the state $|B\rangle$. The explicit form of $|B\rangle$ depends on $\{a_i,c_i\}$. It is summarized in Table~\ref{B_i} (See Appendix B for details).
As is shown in Table~\ref{B_i}, Alice can prepare ten kinds of states, $\{|+_k\rangle\}_{k=0}^7$, $|0\rangle$, and $|1\rangle$ in Bob's place.

Next, we calculate the probability for obtaining each $|+_{k}\rangle$, $|0\rangle$, and $|1\rangle$. The probability that $|B\rangle$ is in the computational basis is
\begin{eqnarray*}
\nonumber
&&{\rm Pr}[|B\rangle=|0\rangle]+{\rm Pr}[|B\rangle=|1\rangle]\\
\nonumber
&=&\cfrac{1}{16}({\rm Pr}[c_3=0]+{\rm Pr}[c_1=0,c_2=c_3=1]\\
\nonumber
&&+{\rm Pr}[c_2=c_5=0,c_3=c_4=1]\\
&&+{\rm Pr}[c_1=c_2=c_3=c_4=1,c_5=0])\\
\nonumber
&=&\cfrac{1}{16}\bigg[p'+\cfrac{1-2p}{1-p}(1-p)(1-p')\\
\nonumber
&&+p(1-p')(1-p)\cfrac{1-2p}{1-p}\\
&&+\cfrac{p}{1-p}(1-p)(1-p')\cfrac{1-2p}{1-p}\bigg]\\
&=&\cfrac{1-4p^2(1-p')}{16}=\cfrac{N_D}{16N}.
\end{eqnarray*}
Since $\{a_i\}$ are chosen uniformly random,
\begin{eqnarray*}
{\rm Pr}[|B\rangle=|0\rangle]={\rm Pr}[|B\rangle=|1\rangle]=\cfrac{N_D}{32N}.
\end{eqnarray*}
By making a similar calculation,
\begin{eqnarray*}
{\rm Pr}[|B\rangle=|+_k\rangle]&=&\cfrac{N-N_D}{16N\times 8}\\
&=&\cfrac{N-N_D}{128N}
\end{eqnarray*}
for each $k\in\{0,\cdot\cdot\cdot,7\}$.
\hspace{\fill}$\blacksquare$
  
\medskip
\section{Blindness}
\label{V}
In this section, we show the blindness of our VBQC protocol.
Remember that, as is explained in Sec.~\ref{III}, our fault-tolerant VBQC protocol is the combination of the gadget (Sec.~\ref{II}) and the FK protocol.
The blindness is shown in three steps.
First, in Sec.~\ref{A}, we introduce a virtual VBQC protocol that is equal to our VBQC protocol except that the gadget of Sec.~\ref{II} is replaced with another ``virtual" gadget.
Second, in Sec.~\ref{B}, we show that the blindness of our VBQC is reduced to that of the virtual VBQC protocol.
In Sec.~\ref{C}, we show the blindness of the virtual VBQC protocol.
As in the previous section, we omit the subscript $L$ of quantum states (e.g. $|A_{i,L}\rangle$) and operators (e.g. $H_L$ and $X_L$) for the notational simplicity.

\subsection{Virtual VBQC protocol}
\label{A}
In this subsection, we explain a virtual VBQC protocol. The virtual VBQC protocol is equivalent to our fault-tolerant VBQC protocol explained in Sec.~\ref{III} except that the gadget is replaced with the following virtual gadget:
\begin{enumerate}
\item Alice sends Bob 
five states 
$\{|A_i\rangle \equiv H^{c_i}X^{a_i}|0\rangle\}_{i=1}^5$ through the quantum channel.
$\{a_i\}_{i=1}^5$ is chosen from $\{0,1\}^{\times5}$ uniformly random.
$c_1$ and $c_5$ are chosen from $\{0,1\}$ with probabilities 
$(1-2p)/(1-p)$ and $p/(1-p)$, respectively.
$c_2$ and $c_4$ are chosen from $\{0,1\}$ with probabilities 
$p$ and $(1-p)$, respectively.
$c_3$ is chosen from $\{0,1\}$ with probabilities $p'$ and $(1-p')$, respectively.
Here, $p$ and $p'$ satisfies that $N_D/N=1-4p^2(1-p')$.

\item Bob performs step $3$ of the (original) gadget explained in Sec.~\ref{II}.
\end{enumerate}
The difference between our gadget in Sec.~\ref{II} and the above virtual gadget is that Alice sends five logical states to Bob, while in our gadget of Sec.~\ref{II} Alice remotely prepares five logical states by measuring halves of logical Bell pairs sent from Bob.

\begin{figure*}[t]
\begin{center}
\includegraphics[width=10cm, clip]{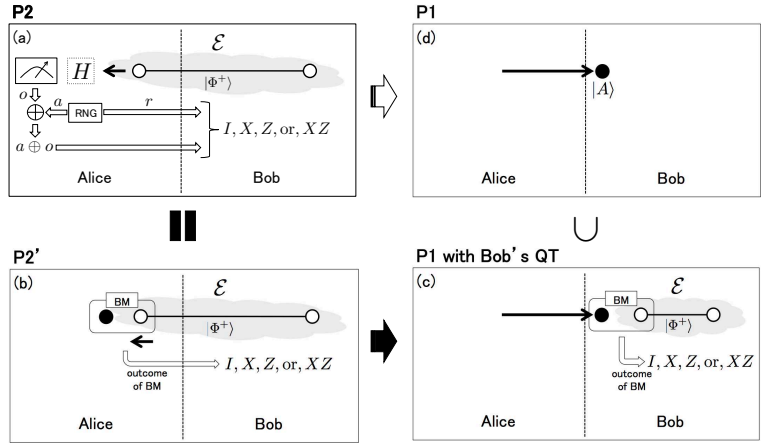}
\end{center}
\captionsetup{justification=raggedright,singlelinecheck=false}
\caption{A diagram of the proof of Theorem \ref{theorem4}. The black and white circles represent the qubits prepared by Alice and Bob, respectively, and $\mathcal{E}$ represents Bob's deviation. For simplicity, we depict only one Bell pair, but, in general, Bob's deviation is applied to all Bell pairs and his ancilla qubits. (a) P2. RNG indicates a random number generator. (b) P2', which is equivalent to P2 from Bob's viewpoint. BM indicates the Bell measurement. (c) BM is delegated to Bob, which only weakens security compared to P2'. (d) P1. (c) can be regarded as a special case of P1.}
\label{equivalence}
\end{figure*}

\subsection{Reduction of our VBQC protocol to the virtual VBQC protocol}
\label{B}
In this subsection, we show that the blindness of our fault-tolerant VBQC protocol in Sec.~\ref{III} can be reduced to that of the virtual VBQC protocol explained in Sec.~\ref{A}. To this end, it is sufficient to show that our gadget given in Sec.~\ref{II} can be reduced to the virtual one given in Sec.~\ref{A}, because other steps of both VBQC protocols are the same.
One might think that this is trivially done by using 
the duality between the state preparation and the measurement 
on a part of a shared entangled state.
However, this is not the case for the following reason: Bob can perform any deviation on the Bell pair $|\Phi^+\rangle$ before sending one half of $|\Phi^+\rangle$ to Alice.
In other words, Bob sends one half of an arbitrary two-qubit state $\rho_{ab}$ instead of one half of $|\Phi^+\rangle$. Here, subscripts $a$ and $b$ represent the system, which is sent to Alice and is kept at Bob's side, respectively. By using Kraus representation, $\rho_{ab}$ can be written as 
\begin{eqnarray*}
\rho_{ab}=\sum_jF_j|\Phi^+\rangle_{ab}\langle\Phi^+|_{ab}F_j^\dag,
\end{eqnarray*}
where $F_j\equiv\langle e_j|_cU_{abc}|e_0\rangle_c$. Here, $|e_j\rangle$ $(0\le j)$ represents an orthonormal basis state of an ancillary system, and $U_{abc}$ represents an unitary operator on the composite system of the systems $a$, $b$, and $c$. By using the property such that 
\begin{eqnarray*}
(I_a\otimes V_b^{\mathrm{T}})|\Phi^+\rangle_{ab}=(V_a\otimes I_b)|\Phi^+\rangle_{ab},
\end{eqnarray*}
$F_j$ can be rewritten as the operator performed on only the system $b$. Here, $V$ represents an unitary operator. In other words, $\rho_{ab}$ can be written as
\begin{eqnarray}
\label{rho_ab}
\rho_{ab}=\mathcal{I}_a\otimes\mathcal{F}_b(|\Phi^+\rangle_{ab}\langle\Phi^+|_{ab}),
\end{eqnarray}
where $\mathcal{F}$ is a super-operator. Since $\mathcal{F}$ is CP (completely-positive) map, but it is not TP (trace-preserving) map in general, we cannot interpret Eq.~(\ref{rho_ab}) such that Bob's deviation, i.e., trace-preserving completely positive (CPTP) map is always performed after sending one half of $|\Phi^+\rangle$ to Alice. 
In fact, when $\rho_{ab}=|+0\rangle_{ab}\langle+0|_{ab}$, 
\begin{eqnarray*}
\mathcal{F}(\cdot)=|0\rangle(\langle 0|+\langle 1|)(\cdot)(|0\rangle+|1\rangle)\langle 0|.
\end{eqnarray*}
Here, $|+\rangle\equiv|+_0\rangle$. In this case, $\mathcal{F}$ is obviously a non-TP map.

However, thanks to random bits used by Alice in our gadget, which acts like twirling~\cite{[BDSW96]}, we can show that 
the duality between the state preparation 
and the measurement holds even under Bob's deviation
as follows:
\begin{theorem} {\bf (Pushing Bob's deviation forward by Alice's randomization)}
\label{theorem4}
Even if Bob sends Alice quantum states different from halves of Bell pairs, what Bob obtains in step 2 of the gadget in Sec.~\ref{II} can be written as $\mathcal{E}(|A\rangle\langle A|)$, where $\mathcal{E}$ is a TPCP map independent on the prepared state $|A\rangle\in\{|+\rangle,|-\rangle,|0\rangle,|1\rangle\}$. Here, $|-\rangle\equiv|+_4\rangle$.
\end{theorem}
{\it Proof.} Hereafter for simplicity, we call the gadget of Sec.~\ref{II} P2. We also call the virtual gadget of Sec.~\ref{A} P1. Furthermore, we define the following modified virtual gadget which we call P2' (See Fig.~\ref{equivalence}):
\begin{enumerate}
\item Bob sends one half of $|\Phi^+\rangle$ through a quantum channel.

\item Alice generates $|A\rangle\in\{|+\rangle,|-\rangle,|0\rangle,|1\rangle\}$ at her side as with P1. In order to send $|A\rangle$ to Bob, Alice performs quantum teleportation (QT) with Bell measurement on $|A\rangle$ and one half of $|\Phi^+\rangle$ sent from Bob.
\end{enumerate}
P2' is equivalent to P2 from Bob's viewpoint (the equivalence between (a) and (b) shown in Fig.~\ref{equivalence}).
The equivalence between them is shown as follows.
Let us consider the Bell measurement on $|A\rangle$ (system 1) and one half of $|\Phi^+\rangle$ (system 2) in QT.
When $|A\rangle$ is a $X$-basis state, it can be written as $X$-bais measurement on one half of $|\Phi^+\rangle$
\begin{widetext}
\begin{eqnarray}
\nonumber
&&{\rm{Tr}}_1\bigg[\cfrac{[I_1I_2+(-1)^{o'}X_1X_2](I_1I_2+Z_1Z_2)}{4}|A\rangle\langle A|\bigg]+{\rm{Tr}}_1\bigg[\cfrac{[I_1I_2+(-1)^{o'}X_1X_2](I_1I_2-Z_1Z_2)}{4}|A\rangle\langle A|\bigg]\\
\nonumber
&=&{\rm{Tr}}_1\bigg[\cfrac{[I_1I_2+(-1)^{o'}X_1X_2](I_1I_2+Z_1Z_2)}{4}\cfrac{I_1+(-1)^{a'}X_1}{2}\bigg]+{\rm{Tr}}_1\bigg[\cfrac{[I_1I_2+(-1)^{o'}X_1X_2](I_1I_2-Z_1Z_2)}{4}\cfrac{I_1+(-1)^{a'}X_1}{2}\bigg]\ \ \ \ \ \ \ \\
\label{QT_X}
&=&2\times\cfrac{I_2+(-1)^{a'\oplus o'}X_2}{4}=\cfrac{I_2+(-1)^{a'\oplus o'}X_2}{2}.
\end{eqnarray}
\end{widetext}
Similarly, when $|A\rangle$ is a $Z$-basis state, it can be written as $Z$-bais measurement on one half of $|\Phi^+\rangle$
\begin{eqnarray}
\nonumber
&&{\rm{Tr}}_1\bigg[\cfrac{(I_1I_2+X_1X_2)[I_1I_2+(-1)^{o'}Z_1Z_2]}{4}\cfrac{I_1+(-1)^{a'}Z_1}{2}\bigg]\\
\nonumber
&+&{\rm{Tr}}_1\bigg[\cfrac{(I_1I_2-X_1X_2)[I_1I_2+(-1)^{o'}Z_1Z_2]}{4}\cfrac{I_1+(-1)^{a'}Z_1}{2}\bigg]\\
\label{QT_Z}
&=&2\times\cfrac{I_2+(-1)^{a'\oplus o'}Z_2}{4}=\cfrac{I_2+(-1)^{a'\oplus o'}Z_2}{2}.
\end{eqnarray}
In P2', two classical bits are sent to Bob per one Bell pair to cancel the byproduct Pauli operator similarly to P2. From Eqs.~(\ref{QT_X}) and (\ref{QT_Z}), a bit corresponding to $r$ is chosen from $\{0,1\}$ with a probability $1/2$, respectively. Accordingly, from Bob's viewpoint, P2 and P2' are completely the same.
Here P2' is further modified in such a way that 
the Bell measurement is delegated to Bob,
which only degrades the blindness and verifiability (the reduction from (b) to (c) shown in Fig.~\ref{equivalence}).
Now, it can be regarded as a special case of P1 (the inclusion of (c) in (d) shown in Fig.~\ref{equivalence}), because in P1, Bob's arbitrary deviation is taken into account.
Accordingly, Bob's deviation in P2 is independent on the prepared four states as with it in P1.
More precisely, from the equivalence between P2 and P2', the state of a qubit prepared at Bob's side after QT can be written as
\begin{widetext}
\begin{eqnarray}
\nonumber
&&\sum_{\tilde{o}_1,\tilde{o}_2}\mathcal{X}_b^{\tilde{o}_1}\mathcal{Z}_b^{\tilde{o}_2}\langle\Phi^+|_{a_1a_2}\mathcal{X}_{a_1}^{\tilde{o}_1}\mathcal{Z}_{a_1}^{\tilde{o}_2}(\rho_{a_1b}\otimes |A\rangle_{a_2}\langle A|_{a_2})|\Phi^+\rangle_{a_1a_2}\\
\nonumber
&=&\sum_{\tilde{o}_1,\tilde{o}_2}\mathcal{X}_b^{\tilde{o}_1}\mathcal{Z}_b^{\tilde{o}_2}\langle\Phi^+|_{a_1a_2}\mathcal{X}_{a_1}^{\tilde{o}_1}\mathcal{Z}_{a_1}^{\tilde{o}_2}(\mathcal{I}_{a_1}\otimes\mathcal{F}_b(|\Phi^+\rangle_{a_1b}\langle\Phi^+|_{a_1b})\otimes |A\rangle_{a_2}\langle A|_{a_2})|\Phi^+\rangle_{a_1a_2}\\
\nonumber
&=&\sum_{\tilde{o}_1,\tilde{o}_2}\mathcal{X}_b^{\tilde{o}_1}\mathcal{Z}_b^{\tilde{o}_2}\langle\Phi^+|_{a_1a_2}\mathcal{I}_{a_1}\otimes\mathcal{F}_b(\mathcal{X}_{a_1}^{\tilde{o}_1}\mathcal{Z}_{a_1}^{\tilde{o}_2}(|\Phi^+\rangle_{a_1b}\langle\Phi^+|_{a_1b}))\otimes |A\rangle_{a_2}\langle A|_{a_2}|\Phi^+\rangle_{a_1a_2}\\
\label{CPTP}
&=&\cfrac{1}{4}\sum_{\tilde{o}_1,\tilde{o}_2}\mathcal{X}_b^{\tilde{o}_1}\mathcal{Z}_b^{\tilde{o}_2}\mathcal{F}_b\mathcal{X}_b^{\tilde{o}_1}\mathcal{Z}_b^{\tilde{o}_2}(|A\rangle_b\langle A|_b).
\end{eqnarray}
\end{widetext}
Here, $\mathcal{X}^{\tilde{o}_1}(\cdot)\equiv X^{\tilde{o}_1}(\cdot)X^{\tilde{o}_1}$ and $\mathcal{Z}^{\tilde{o}_2}(\cdot)\equiv Z^{\tilde{o}_2}(\cdot)Z^{\tilde{o}_2}$, where $(\tilde{o}_1,\tilde{o}_2)\in\{0,1\}^{\times 2}$.
Since $\mathcal{F}_b$ is CP map, $1/4\sum_{\tilde{o}_1,\tilde{o}_2}\mathcal{X}_b^{\tilde{o}_1}\mathcal{Z}_b^{\tilde{o}_2}\mathcal{F}_b\mathcal{X}_b^{\tilde{o}_1}\mathcal{Z}_b^{\tilde{o}_2}$ is also CP map. Next, we show that $1/4\sum_{\tilde{o}_1,\tilde{o}_2}\mathcal{X}_b^{\tilde{o}_1}\mathcal{Z}_b^{\tilde{o}_2}\mathcal{F}_b\mathcal{X}_b^{\tilde{o}_1}\mathcal{Z}_b^{\tilde{o}_2}$ is TP map. Let $|\psi\rangle$ be a single-qubit state. By using Eq.~(\ref{rho_ab}),
\begin{eqnarray}
\nonumber
&&{\rm Tr}\left[\cfrac{1}{4}\sum_{\tilde{o}_1,\tilde{o}_2}\mathcal{X}_b^{\tilde{o}_1}\mathcal{Z}_b^{\tilde{o}_2}\mathcal{F}_b\mathcal{X}_b^{\tilde{o}_1}\mathcal{Z}_b^{\tilde{o}_2}(|\psi\rangle_b\langle\psi|_b)\right]\\
\nonumber
&=&\cfrac{1}{4}\sum_{\tilde{o}_1,\tilde{o}_2}{\rm Tr}[\mathcal{F}_b\mathcal{X}_b^{\tilde{o}_1}\mathcal{Z}_b^{\tilde{o}_2}(|\psi\rangle_b\langle\psi|_b)]\\
\label{TP}
&=&{\rm Tr}\left[\mathcal{F}_b\left(\cfrac{I_b}{2}\right)\right]={\rm Tr}[\rho_b]=1.
\end{eqnarray}
From Eq.~(\ref{TP}), unlike Eq.~(\ref{rho_ab}), Eq.~(\ref{CPTP}) can be interpreted such that TPCP map, which is independent of $|A\rangle$, is applied for a qubit prepared by Alice as Bob's deviation similar to P1.
 \hspace{\fill}$\blacksquare$

\subsection{Blindness of the virtual VBQC protocol}
\label{C}

Let $\rho_{ab}$ is the output of the virtual gadget where subscripts $a$ and $b$ denote Alice's and Bob's systems, respectively, and Alice's classical registers are treated as quantum states. If Bob is malicious and did not follow the correct procedure, $\rho_{ab}$ can be any state. We define $\rho_{ab}^{({\rm FK})}$, which is a state prepared in the state-preparation step of the (original) FK protocol~\cite{[FK12]}, by
\begin{eqnarray*}
&&\rho_{ab}^{({\rm FK})}\\
&\equiv&\mathcal{E}_b\left(\sum_{z=0}^1P[\sqrt{p(z)}|z\rangle_a|z\rangle_b]+\sum_{k=0}^7P[\sqrt{p(k)}|k\rangle_a|+_k\rangle_b]\right),
\end{eqnarray*}
where $P[|\cdot\rangle]\equiv|\cdot\rangle\langle\cdot |$, and $\mathcal{E}_b$ represents Bob's deviation (TPCP map).
Here, as mentioned earlier, subscripts $a$ and $b$ denote Alice's and Bob's systems, respectively, and Alice's classical registers are treated as quantum states.
Finally, let $\Pi_b$ be any positive operator valued measure (POVM) element performed on Bob's system.
In order to show the blindness of the virtual VBQC protocol, it is sufficient to show
\begin{eqnarray}
\label{blindness}
{\rm Tr}[\Pi_b\rho_{ab}]={\rm Tr}[\Pi_b\rho_{ab}^{({\rm FK})}].
\end{eqnarray}

For the virtual gadget, the following lemma holds:
\begin{lemma}
\label{theorem2}
If Alice follows the procedure of the virtual gadget, its output state satisfies Eq.~(\ref{blindness}) for any POVM element performed on Bob's system.
\end{lemma}
{\it{Proof.}} We define $U_b$ as the unitary operator performed in Fig.~\ref{mds_graph}. Without loss of generality, we can assume that Bob performs deviation and projection to the case where $s_1=s_2=s_4=s_5=0$ after performing $U_b$ (See Appendix C or \cite{[FK12]} for the reason). We define CP map $\mathcal{E'}_b$ as such operation. Before $Z$-basis measurements in Fig.~\ref{mds_graph}, a state that is composed of Alice's registers and five qubits at Bob's side can be written as
\begin{eqnarray*}
\rho_{ab}\propto\mathcal{E'}_b\left(U_b\bigotimes_{i=1}^5P\left[\sqrt{p(a_i)p(c_i)}|a_ic_i\rangle_{a^{(i)}}|A_i\rangle_{b^{(i)}}\right]U_b^\dag\right),
\end{eqnarray*}
where Alice's system $a$ and Bob's system $b$ are composed of systems $\{a^{(i)}\}_{i=1}^5$ and $\{b^{(i)}\}_{i=1}^5$, respectively.
Let $\rho_{ab^{(3)}}\equiv{\rm Tr}_{\tilde{b}}[\rho_{ab}]$ be a reduced density operator obtained by taking the partial trace over systems $\tilde{b}\equiv\{b^{(1)},b^{(2)},b^{(4)},b^{(5)}\}$.
From Table~\ref{B_i} and taking the case where at least one of $\{s_1,s_2,s_4,s_5\}$ is not equal to $0$ into account, it can be calculated as
\begin{widetext}
\begin{eqnarray}
\nonumber
\rho_{ab^{(3)}}&=&\cfrac{1}{P}{\rm Tr}_{\tilde{b}}\left[\mathcal{E'}_b\left(\cfrac{1}{16}\sum_{{\bf s},{\bf s'}\in\{0,1\}^{\times 4}}\mathcal{E'}_a^{{\bf s},{\bf s'}}\left(\sum_{z=0}^1P[\sqrt{p(z)}|z\rangle_a|z\rangle_{b^{(3)}}]+\sum_{k=0}^7P[\sqrt{p(k)}|k\rangle_a|+_k\rangle_{b^{(3)}}]\right)\otimes|s_1s_2s_4s_5\rangle\langle s'_1s'_2s'_4s'_5|_{\tilde{b}}\right)\right]\\
\label{8}
&\equiv&\tilde{\mathcal{E}}_a\left(\rho_{ab^{(3)}}^{({\rm FK})}\right),
\end{eqnarray}
\end{widetext}
where $P$ is a probability where $s_1=s_2=s_4=s_5=0$ is obtained, $\tilde{\mathcal{E}}_a$ is a TPCP map performed on Alice's system, and $\mathcal{E'}_a^{{\bf s},{\bf s'}}$ is an operation performed on Alice's system depending on ${\bf s}\equiv\{s_1,s_2,s_4,s_5\}$ and ${\bf s'}\equiv\{s'_1,s'_2,s'_4,s'_5\}$. Note that ${\bf s}$ and ${\bf s'}$ are independent of the form of $\rho^{({\rm FK})}$. As shown in Theorem~\ref{theorem1_2}, $\mathcal{E'}_a^{{\bf 0},{\bf 0}}=\mathcal{I}_a$, where $\mathcal{I}$ is the identity super-operator. Accordingly, 
\begin{eqnarray*}
{\rm Tr}[\Pi_{b^{(3)}}\rho_{ab^{(3)}}]={\rm Tr}[\Pi_{b^{(3)}}\rho_{ab^{(3)}}^{({\rm FK})}]
\end{eqnarray*}
is satisfied for any Bob's POVM element $\Pi_{b^{(3)}}$. This means that if the virtual gadget is used as the state-preparation step of the FK protocol, it does not degrade blindness.
\hspace{\fill}$\blacksquare$

From Lemma~\ref{theorem2} and Theorem~\ref{theorem4}, the following theorem immediately holds:
\begin{theorem}
\label{theorem2_2}
Our fault-tolerant VBQC protocol satisifies the blindness.
\end{theorem}
Note that although we consider only single run of our gadget in above proofs, the similar argument also holds when Bob performs deviation on all of logical Bell pairs used in multiple run of our gadget.

\medskip
\section{Verifiability}
\label{VI}
In this section, we show that our fault-tolerant VBQC protocol satisfies the verifiability. As in the previous section, we first show the verifiability of the virtual VBQC protocol, and then we reduce the verifiability of our fault-tolerant VBQC protocol to that of the virtual one. Again, we omit the subscript $L$ of quantum states and operators for the notational simplicity.

For the virtual VBQC protocol, following lemma holds:
\begin{lemma}
\label{theorem3}
The virtual VBQC protocol satisfies the verifiability.
\end{lemma}
{\it Proof.} A detailed proof is given in Appendix C. Here, we explain intuitive ideas for the proof.
Our proof is similar to that of the verifiability of the original FK protocol~\cite{[FK12]}.
Hereafter, we briefly explain why the proof of the original FK protocol is used to show Lemma~\ref{theorem3}. The virtual gadget in Sec.~\ref{V} satisfies following two properties:
\begin{remark}
(i) When Alice and Bob follow the correct procedure of the virtual gadget, an output state $\rho_{ab}$ that represents classical-quantum correlation between Alice and Bob satisfies that $\rho_b=(I/2)^{\otimes{\rm log}({\rm dim}\rho_b)}$. Here, $\rho_b$ and ${\rm dim}\rho_b$ represent the reduced density operator for Bob's system and dimension of Bob's system, respectively.
(ii) Bob's deviation is independent of the states prepared by Alice.
\end{remark}
The first property is derived from the fact that when Bob is honest, $\mathcal{E}_b=\mathcal{I}_b$ in Eq.~(\ref{8}). Furthermore, since $\mathcal{E}_b$ is independent of $\{|A_i\rangle\}$, the second property is also satisfied. Note that our gadget can be treated as a special case of the virtual one, these two properties are also satisfied for the gadget in Sec.~\ref{II}.

These two properties are sufficient conditions to utilize techniques used in proof of verifiability of the FK protocol~\cite{[FK12]}. Accordingly, they are important to show that our gadget does not degrade verifiability of the FK protocol. Note that property (i) is not always necessary for blindness. In fact, we do not use property (i) to show blindness. 
The reason why these two properties are required is as follows: For the FK protocol, an average probability where Alice accepts an incorrect outcome over her secret information is calculated to show verifiability. Here, we define $\nu$, $\rho(\nu)$, $\mathcal{T}$, $\mathcal{W}$, and $\Pi$ as Alice's secret information, an initial state prepared in Bob's place, an ideal operation performed by Alice and honest Bob, Bob's deviation, and a projector composed of a projector performed in the FK protocol and a projector corresponding to the event where Alice accepts an incorrect outcome, respectively. In the FK protocol, it is assumed that $\mathcal{W}$ is independent of $\nu$. In order to satisfy this assumption for our gadget, we require property (ii). Since $\mathcal{W}$ can be decomposed by multi-qubit Pauli operators, in order to calculate the average probability, we have to calculate
\begin{eqnarray}
\label{pro1}
\sum_\nu p(\nu){\rm Tr}\left[\Pi\sigma\mathcal{T}(\rho(\nu))\sigma'\right]
\end{eqnarray}
for several $\sigma$ and $\sigma'$, where $\sigma$ and $\sigma'$ are multi-qubit Pauli operators, and $p(\nu)$ is a probability where Alice selects $\nu$. Note that we can assume that Bob's deviation is performed after the ideal operation without loss of generality as shown in \cite{[FK12]} and Appendix C. In the FK protocol, 
\begin{eqnarray*}
\sum_{\nu}p(\nu)\rho(\nu)=\left(\cfrac{I}{2}\right)^{\otimes{\rm log}({\rm dim}\rho(\nu))}
\end{eqnarray*}
is satisfied and then Eq.~(\ref{pro1}) becomes $0$ when $\sigma\neq\sigma'$. This fact is important to complete the proof, and we require property (i) to use this fact in our proof of verifiability (See Appendix C for a detailed proof).

As an example that does not satisfy (i), in Fig.~\ref{mds_graph}, we can 
replace $T|A_1\rangle$ and $S|A_5\rangle$ with $|+_1\rangle$ and $|+_2\rangle$, respectively.
Let Bob then prepare $|+_1\rangle$ and $|+_2\rangle$ at Bob's side, 
similarly to Ref.~\cite{[DK16]}.
In this example,  
the correctness and blindness are satisfied.
However, because Bob's initial states are not the maximally mixed state from Bob's viewpoint even in the ideal case,
the verifiability cannot be guaranteed by using the same argument in Ref.~\cite{[FK12]}.
As another example that does not satisfy (ii), we can remove the discarding procedure in our gadget. Even though, the correctness and blindness are satisfied similar to the above example, and the success probability is increased to 1. However, since the prepared state depends on $\{s_i\}$, Bob can perform deviation depending on the state prepared by Alice. To avoid such a situation, the discarding procedure is required.
\hspace{\fill}$\blacksquare$

From Theorem~\ref{theorem4} and Lemma~\ref{theorem3}, the following theorem immediately holds:
\begin{theorem}
\label{theorem3_2}
Our fault-tolerant VBQC protocol satisfies the verifiability.
\end{theorem}

\medskip
\begin{center}
{\bf ACKNOWLEDGMENTS}
\end{center}

We thank Masato Koashi and Akihiro Mizutani for helpful discussions. YT is supported by Program for Leading Graduate Schools: “Interactive Materials Science Cadet Program” and JSPS Grant-in-Aid for JSPS Research Fellow No.JP17J03503.
KF is supported by KAKENHI No. 16H02211, JST PRESTO JPMJPR1668, JST ERATO JPMJER1601, and JST CREST JPMJCR1673.
TM is supported by JST ACT-I No.JPMJPR16UP, the Grant-in-Aid for Scientific Research on Innovative Areas No.15H00850 of MEXT Japan, the JSPS Grant-in-Aid for Young Scientists (B) No.26730003 and No.17K12637, and JST, PRESTO.
NI is supported by JSPS KAKENHI Grant No. JP16H02214 and JST CREST JPMJCR1671.

\medskip
\begin{center}
{\bf APPENDIX A: THE FK PROTOCOL}
\end{center}

In this appendix, we briefly explain the procedure of the FK protocol~\cite{[FK12]}. The FK protocol runs as follows:
\begin{enumerate}
\item Alice prepares a qubit, and sends it to Bob through a quantum channel. 
Alice repeats this procedure $N$ times. $N_D$ of $N$ qubits are each of which chosen from the $Z$-basis states uniformly random. We call these qubits dummy qubits. $(N-N_D)$ qubits are each chosen from $\{|+_k\rangle\}_{k=0}^7$ uniformly random.

\item Bob generates a randomly-rotated dotted-complete graph state 
by entangling $N$ qubits sent from Alice according to Alice's instruction. The randomly-rotated dotted-complete graph state $|{\rm RDC}\rangle$ is defined as 
\begin{eqnarray*}
\prod_{(i,j)\in E}\Lambda_{i,j}(Z)\left(\prod_{\tilde{i}=1}^{N-N_D}|+_{k_{\tilde{i}}}\rangle_{\tilde{i}}\prod_{\tilde{i}=N-N_D+1}^N|z_{\tilde{i}}\rangle_{\tilde{i}}\right).
\end{eqnarray*}
Here, $E$ is defined as a set of edges of a dotted-complete graph introduced in Ref.~\cite{[FK12]}, $|+_{k_{\tilde{i}}}\rangle_{\tilde{i}}\equiv(|0\rangle_{\tilde{i}}+e^{ik_{\tilde{i}}\pi/4}|1\rangle_{\tilde{i}})/\sqrt{2}$, and $|z_{\tilde{i'}}\rangle_{\tilde{i'}}$ $(z_{\tilde{i'}}\in\{0,1\})$ is the $\tilde{i'}$th $Z$-basis state.

\item Alice sends a value of $\delta_{i'}\equiv {{k'}_{i'}\pi}/4+\phi_{i'}+{r'}_{i'}\pi+n_{i'}\pi$ to Bob through a classical channel, then Bob measures the $i'$th qubit $(1\le i'\le N, i'\in\mathbb{N})$ of $|{\rm RDC}\rangle$ in $\{|+_{\delta_{i'}4/\pi}\rangle,|+_{4+\delta_{i'}4/\pi}\rangle\}$, 
and sends the outcome $b_{i'}$ to Alice through the classical channel.
For any qubits, $r_{i'}$ is chosen from $\{0,1\}$ uniformly random. 
$n_{i'}$ is the number of $|1\rangle$, 
which are neighbors of the $i'$th qubit on $|{\rm RDC}\rangle$. To remove the effect of $Z_{i'}^{n_{i'}}$, the term $n_{i'}\pi$ is necessary. In Ref.~\cite{[FK12]}, 
the effect of the term $n_{i'}\pi$ is considered in step 1, 
but in this paper it is consider in step 3 to make the FK protocol appropriately for our gadget. This modification does not lose the essential properties of the FK protocol at all.
For each of the dummy qubits, 
the value of ${k'}_{i'}$ is choosen from $\{0,1,2,3,4,5,6,7\}$ uniformly random. 
For other qubits whose state is $|+_{k_{\tilde{i}}}\rangle$, ${k'}_{i'}=k_{\tilde{i}}$. 
For dummy qubits, $\phi_{i'}(\in\{k\pi/4\}_{k=0}^7)$ is chosen uniformly random. For other qubits used to perform universal quantum computing, $\phi_{i'}$ is chosen according to the quantum algorithm where Alice wants to perform and previous measurement outcomes as with MBQC. For other qubits used to perform the verification, i.e., trap qubits, $\phi_{i'}$ is chosen as $0$.

\item Alice checks whether or not $b_{i'}=r_{i'}$ is satisfied for all trap qubits. 
If it is satisfied, 
Alice accepts the output of her desired quantum computing. 
Otherwise, Alice rejects it.
\end{enumerate}

\medskip
\begin{center}
{\bf APPENDIX B: THE PROOF FOR CORRECTNESS OF OUR FAULT-TOLERANT VBQC PROTOCOL}
\end{center}

In this appendix, we derive Table~\ref{B_i}. Note that we omit the subscript $L$ of quantum states and operators for the notational simplicity.

First, we consider step $2$. If Alice measures one half of $|\Phi^+\rangle$ in $Z$ basis and obtains the measurement outcome $o_i$, $X^{a_i\oplus o_i}Z^{r_i}|o_i\rangle=(-1)^{r_i\cdot o_i}|a_i\rangle$ is prepared at Bob's side. On the other hand, if Alice measures one half of $|\Phi^+\rangle$ in $X$ basis and obtains the measurement outcome $o_i$, $X^{r_i}Z^{a_i\oplus o_i}|+_{4o_i}\rangle=(-1)^{r_i\cdot a_i}|+_{4a_i}\rangle$ is prepared at Bob's side. Hence, $|+_0\rangle$, $|+_4\rangle$, $|0\rangle$, and $|1\rangle$ are prepared at Bob's side with probabilities $q_i/2$, $q_i/2$, $(1-q_i)/2$, and $(1-q_i)/2$, respectively.

Next, we consider step $3$. Here, we consider only the case of $s_i=0$ $(i=1,2,4,5)$ because in other cases, Bob discards $|B\rangle$. The probability that $s_1=s_2=s_4=s_5=0$ is satisfied is $1/16$ independent of the form of $|B\rangle$.
From a calculation by taking into account the dependence of $|B\rangle$ on $\{a_i\}$ and $\{c_i\}$, $|B\rangle$ is derived as shown in Table~\ref{B_i}.
Below we will explain how the calculation proceeds.
When $(c_2,c_3,c_4)=(0,1,0)$, 
$|B\rangle$ is an eigenstate of $X$ because the 3rd qubit is not connected to other four qubits. 
Similarly, when $c_3=0$, the $3$rd qubit is not connected to other four qubits, 
and so $|B\rangle$ is an eigenstate of $Z$. 
When $|A_1\rangle$ is connected to the 3rd qubit 
through $|A_2\rangle$ ($c_1=c_2=c_3=1$), 
by measuring $T|A_1\rangle$ and $|A_2\rangle$ in $X$ bases, 
$T$ or $T^\dag$ is performed on the $3$rd qubit up to the byproduct operators via gate teleportation. 
On the other hand, when $c_3=c_4=c_5=1$, $S$ is performed on 
the 3rd qubit in a similar way. 
When $(c_1,c_2,c_3)=(0,1,1)$ or $(c_3,c_4,c_5)=(1,1,0)$, $H$ is performed on the $3$rd qubit in the similar way, 
therefore $|B\rangle$ is an eigenstate of $Z$. 
From the above observation, 
Alice can prepare $|B\rangle$ up to a global phase as in Table~\ref{B_i}.

\medskip
\begin{center}
{\bf APPENDIX C: THE PROOF FOR VERIFIABILITY OF THE VIRTUAL VBQC PROTOCOL}
\end{center}

We employ almost the same method used in Ref.~\cite{[FK12]}. Note that we omit the subscript $L$ of quantum states and operators for the notational simplicity.

\begin{figure*}[t]
\begin{center}
\includegraphics[width=15cm, clip]{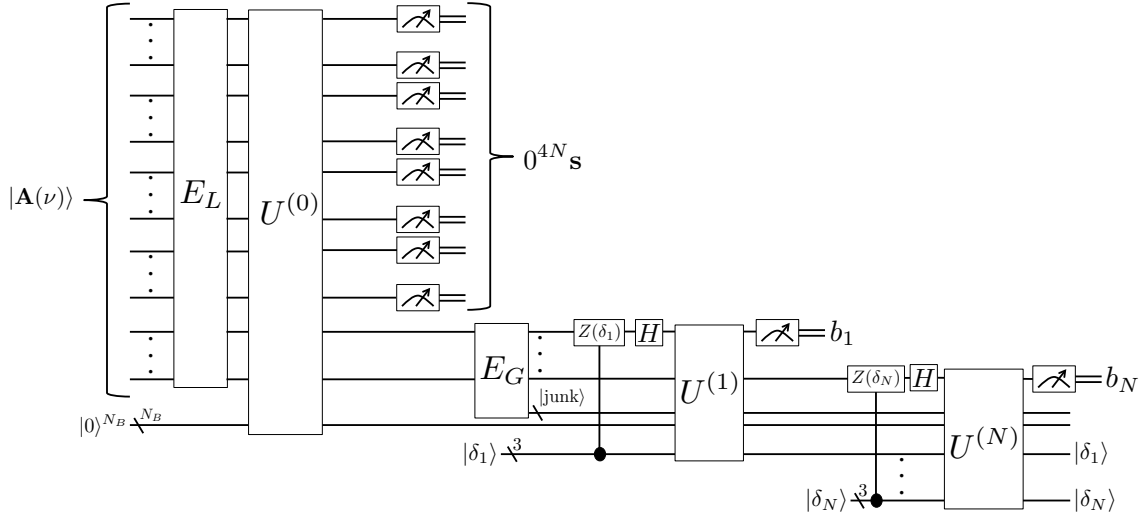}
\end{center}
\captionsetup{justification=raggedright,singlelinecheck=false}
\caption{A circuit diagram of our fault-tolerant VBQC protocol including Bob's deviation $U^{(i')}$. 
The classical message $\delta_{i'}$ is denoted as three-qubit quantum state $|\delta_{i'}\rangle$, and ${\bf s}$ represents the outcomes in step 2 of the virtual gadget that decide the states of discarded qubits.
The detail of the notations is written in the main text.}
\label{verification}
\end{figure*}

\begin{figure*}[t]
\begin{center}
\includegraphics[width=12cm, clip]{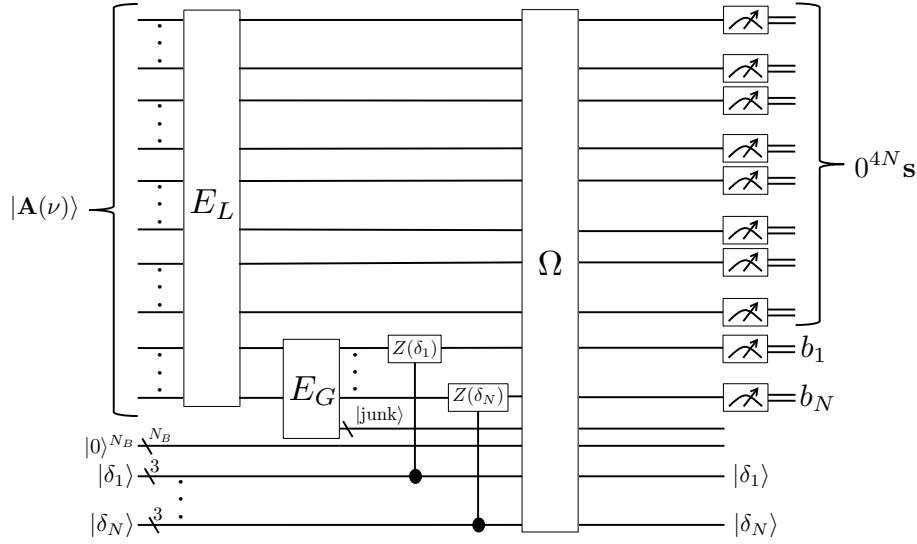}
\end{center}
\captionsetup{justification=raggedright,singlelinecheck=false}
\caption{A modified circuit diagram of our fault-tolerant VBQC protocol including postponed Bob's deviation $\Omega$.}
\label{verification_transform}
\end{figure*}

A circuit diagram of our fault-tolerant VBQC protocol is shown in Fig.~\ref{verification}. Bob's $(i''+1)$th deviation is denoted by $U^{(i'')}$ $(0\le i''\le N)$. Particularly, the deviations performed in the virtual gadget are included in $U^{(0)}$.
In Fig.~\ref{verification}, 
\begin{eqnarray*}
|{\bf A}(\nu)\rangle\equiv\bigotimes_{j=1}^{N'}(|A_{5j-4}\rangle|A_{5j-3}\rangle|A_{5j-2}\rangle|A_{5j-1}\rangle|A_{5j}\rangle), 
\end{eqnarray*}
$j$ means the $j$th repetition of the virtual gadget, 
$E_{L}$ represents Bob's faithful operation before the $Z$-basis measurements shown in Fig.~\ref{mds_graph}, 
\begin{eqnarray*}
E_{G}(\nu)\equiv \left(\prod_{(i,j)\in E}\Lambda_{i,j}(Z)\right)\otimes I^{\otimes N'-N},
\end{eqnarray*}
$|{\rm junk}\rangle$ represents the discarded qubits,
and $|0\rangle^{\otimes N_B}$ is the ancilla qubits, 
which are used to make Bob's deviation unitary operators. 
Here, 
Alice's random variable $\nu$ represents the random value $r_{i'}$ mentioned in step 3 of the FK protocol, ${\bf a}^{(j)}\equiv\{a_{5j-4},a_{5j-3},a_{5j-2},a_{5j-1},a_{5j}\}$, and ${\bf c}^{(j)}\equiv\{c_{5j-4},c_{5j-3},c_{5j-2},c_{5j-1},c_{5j}\}$. 
Note that $\tilde{N}$ of $N$ outcomes $\{b_{i'}\}$ represent the output of Alice's delegated quantum computing. 
In this proof, 
we denote the classical bits as quantum states such as $\delta_{i'}\rightarrow|\delta_{i'}\rangle$. It is known that Bob's deviation $U_{i'}$ does not depend on $\nu$ (property (ii) in Remark 1). 
In order to calculate the probability of Alice accepting the incorrect output, 
we postpone Bob's deviation depicted in Fig.~\ref{verification} without changing quantum states just before measurements as 
shown in Fig.~\ref{verification_transform}.
Now, we define that 
\begin{eqnarray*}
T&\equiv& \left(\prod_{i'=1}^NH_{i'}Z_{i'}(\delta_{i'})\right)E_GE_L, \\
T^{(0)}&\equiv& TE_L^\dag, \\
T^{(i')}&\equiv&\prod_{j'=i'+1}^NH_{j'}Z_{j'}(\delta_{j'}), \\
\Omega&\equiv&\prod_{i''=0}^NT^{(i'')}U^{(i'')}{T^{(i'')}}^\dag, \\
|\Psi(\nu)\rangle&\equiv&|{\bf A}(\nu)\rangle\left(\bigotimes_{i'=1}^N|\delta_{i'}\rangle\right). 
\end{eqnarray*}
Here, $Z_{i'}(\delta_{i'})\equiv |0\rangle\langle 0|_{i'}+e^{-i\delta_{i'}}|1\rangle\langle 1|_{i'}$, and $\Omega$ represents the postponed Bob's deviation. Note that if Bob is honest, $\Omega$ is the identity operator. Moreover, for simplicity, we define 
\begin{eqnarray*}
\mathcal{T}(\cdot)&\equiv& T(\cdot)T^\dag, \\
\mathcal{W}(\cdot)&\equiv& \Omega(\cdot)\Omega^\dag.
\end{eqnarray*}
The output quantum state $A(\nu)$ composed of all qubits except for ancilla qubits can be written as 
\begin{widetext}
\begin{eqnarray*}
A(\nu)=\cfrac{1}{p_n}{\rm Tr}_B\Bigg[\sum_{{\bf s},{\bf b}}P[|0\rangle^{\otimes 4N}|{\bf s}\rangle]|{\bf b'}\rangle\langle {\bf b}|\mathcal{W}\mathcal{T}(P[|\Psi(\nu)\rangle]\otimes P[|0\rangle^{\otimes N_B}])|{\bf b}\rangle\langle {\bf b'}|\Bigg].
\end{eqnarray*}
\end{widetext}
Here, $p_n$ is the normalization factor, $|{\bf s}\rangle$ represents the outcomes in step 2 of the virtual gadget that decide the states of discarded qubits, the state
\begin{eqnarray*}
|{\bf b}\rangle\equiv\prod_{\tilde{j}=1}^{N-\tilde{N}}|b_{\tilde{j}}\rangle_{\tilde{j}}
\end{eqnarray*}
represents the outcomes that are not output of Alice's delegated quantum computing, 
\begin{eqnarray*}
|{\bf b'}\rangle\equiv\prod_{\tilde{j}=1}^{N-\tilde{N}}|b_{\tilde{j}}\oplus r_{\tilde{j}}\rangle_{\tilde{j}}, 
\end{eqnarray*}
and ${\rm Tr}_B[\cdot]$ represents the partial trace over Bob's ancilla qubits. 
Next, we define a projector onto the subspace spanned by the states of the non-trap qubits used in the FK protocol that generates incorrect output as $\Pi_{\perp}$, and define the set of positions of the trap qubits as $T'(\nu)$, 
respectively. 
The probability ${p'}_{\rm incorrect}$ where Alice accepts an incorrect output is calculated to be
\begin{widetext}
\begin{eqnarray*}
\nonumber
&&{p'}_{\rm incorrect}\\
&=&\sum_{\nu}p(\nu){\rm Tr}[\Pi_{\perp}P[\otimes_{t\in T'(\nu)}|r_t\rangle]A(\nu)]\\
&=&\sum_{\nu}p(\nu){\rm Tr}\Bigg[\Pi_{\perp}P[\otimes_{t\in T'(\nu)}|r_t\rangle]\Bigg(\cfrac{1}{p_n}\sum_{{\bf s},{\bf b}}P[|0\rangle^{\otimes 4N}|{\bf s}\rangle]|{\bf b'}\rangle\langle {\bf b}|\mathcal{W} \mathcal{T}(P[|\Psi(\nu)\rangle]\otimes P[|0\rangle^{\otimes N_B}])|{\bf b}\rangle\langle {\bf b'}|\Bigg)\Bigg]\ \ \ \ \ \ \\
&\equiv&\cfrac{p_{\rm incorrect}}{p_n}.
\end{eqnarray*}
\end{widetext}
We define a Kraus operator $\chi_{k'}\equiv\langle k'|\Omega|0\rangle^{\otimes N_B}$, where $\{|k\rangle\}$ are the normal orthogonal bases for the Hilbert space corresponding to the input state of Fig.~\ref{verification_transform} except Bob's ancilla qubits. From this definition, 
\begin{widetext}
\begin{eqnarray*}
\nonumber
&&p_{\rm incorrect}\\
&=&\sum_{\nu}p(\nu){\rm Tr}\Bigg[\Pi_{\perp}P[\otimes_{t\in T'(\nu)}|r_t\rangle]\Bigg(\sum_{{\bf s},{\bf b},k'}P[|0\rangle^{\otimes 4N}|{\bf s}\rangle]|{\bf b'}\rangle\langle {\bf b}|\chi_{k'}\mathcal{T} (P[|\Psi(\nu)\rangle])\chi_{k'}^\dag|{\bf b}\rangle\langle {\bf b'}|\Bigg)\Bigg].
\end{eqnarray*}
\end{widetext}
Since the Kraus operator can be written as a liner combination of the tensor products $\{\sigma_{\tilde{j'}}\}$ of Pauli operators with complex coefficients, 
$\chi_{k'}=\sum_{\tilde{j'}}\alpha_{k'\tilde{j'}}\sigma_{\tilde{j'}}$, where $\sum_{k',\tilde{j'}}|\alpha_{k'\tilde{j'}}|^2=1$, 
is satisfied. Accordingly, 
\begin{widetext}
\begin{eqnarray*}
\nonumber
&&p_{\rm incorrect}\\
&=&\sum_{\nu}p(\nu){\rm Tr}\Bigg[\Pi_{\perp}P[\otimes_{t\in T'(\nu)}|r_t\rangle]\Bigg(\sum_{{\bf s},{\bf b},k',\tilde{j'},j''}\alpha_{k'\tilde{j'}}\alpha_{k'j''}^\ast P[|0\rangle^{\otimes 4N}|{\bf s}\rangle]|{\bf b'}\rangle\langle {\bf b}|\sigma_{\tilde{j'}}\mathcal{T}(P[|\Psi(\nu)\rangle])\sigma_{j''}|{\bf b}\rangle\langle {\bf b'}|\Bigg)\Bigg]\ \ \ \ \ \ \ \\
&=&\sum_{\nu,{\bf s},{\bf b},k'}p(\nu){\rm Tr}\Bigg[\Pi_{\perp}P[\otimes_{t\in T'(\nu)}|r_t\rangle]\Bigg(\sum_{\tilde{j'}}\sum_{j''}\alpha_{k'\tilde{j'}}\alpha_{k'j''}^\ast P[|0\rangle^{\otimes 4N}|{\bf s}\rangle]|{\bf b'}\rangle\langle {\bf b}|\sigma_{\tilde{j'}}\mathcal{T}(P[|\Psi(\nu)\rangle])\sigma_{j''}|{\bf b}\rangle\langle {\bf b'}|\Bigg)\Bigg]\\
&=&\sum_{\nu,{\bf s},{\bf b},k'}p(\nu){\rm Tr}\Bigg[\Pi_{\perp}P[\otimes_{t\in T'(\nu)}|r_t\rangle]\Bigg[\sum_{\tilde{j'}}\sum_{j''}\alpha_{k'\tilde{j'}}\alpha_{k'j''}^\ast P[|0\rangle^{\otimes 4N}|{\bf s}\rangle](\otimes_t|r_t\rangle)\langle {\bf b}|\sigma_{\tilde{j'}}\mathcal{T}(P[|\Psi(\nu)\rangle])\sigma_{j''}|{\bf b}\rangle(\otimes_t\langle r_t|)\Bigg]\Bigg]\ \ \ \ \ \ \ \\
&=&\sum_{\nu,{\bf s},{\bf b},k'}p(\nu){\rm Tr}\Bigg[\Pi_{\perp}P[\otimes_{t\in T'(\nu)}|r_t\rangle]\Bigg(\sum_{\tilde{j'}}\sum_{j''}\alpha_{k'\tilde{j'}}\alpha_{k'j''}^\ast P[|0\rangle^{\otimes 4N}|{\bf s}\rangle]\langle {\bf b}|\sigma_{\tilde{j'}}\mathcal{T}(P[|\Psi(\nu)\rangle])\sigma_{j''}|{\bf b}\rangle\Bigg)\Bigg]\\
&=&\sum_{\nu,{\bf s},{\bf b'},k'}p(\nu){\rm Tr}\Bigg[\Pi_{\perp}P[\otimes_{t\in T'(\nu)}|r_t\rangle]\Bigg(\sum_{\tilde{j'}}\sum_{j''}\alpha_{k'\tilde{j'}}\alpha_{k'j''}^\ast P[|0\rangle^{\otimes 4N}|{\bf s}\rangle|{\bf b'}\rangle]\sigma_{\tilde{j'}}\mathcal{T}(P[|\Psi(\nu)\rangle])\sigma_{j''}\Bigg)\Bigg]\\
&\le&\sum_{\nu,{\bf s},{\bf b'},k'}p(\nu){\rm Tr}\Bigg[P[\otimes_{t\in T'(\nu)}|r_t\rangle]\Bigg(\sum_{\tilde{j'}}\sum_{j''}\alpha_{k'\tilde{j'}}\alpha_{k'j''}^\ast P[|0\rangle^{\otimes 4N}|{\bf s}\rangle|{\bf b'}\rangle]\sigma_{\tilde{j'}}\mathcal{T}(P[|\Psi(\nu)\rangle])\sigma_{j''}\Bigg)\Bigg].
\end{eqnarray*}
\end{widetext}
Here, ${\bf b'}\equiv\{b_{\tilde{j}}|\tilde{j}\neq t\}$. 
We divide $\nu$ into $\nu_T$ and its complementary set $\bar{\nu}_T$, 
where 
$\nu_T$ represents the position of the trap qubits, $\{{\bf a}^{(t)}\}$, $\{{\bf c}^{(t)}\}$, and $\{r_t\}$. 
Since 
\begin{eqnarray*}
&&\sum_{\bar{\nu}_T}p(\bar{\nu}_T)\mathcal{T}(P[|\Psi(\nu)\rangle])\\
&=&\otimes_tP[\sum_{{\bf s}_T}\sqrt{p({\bf s}_T)}|{\bf s}_T\rangle HZ(\delta_t)|B_t\rangle|\delta_t\rangle]\otimes(I/2)^{\otimes\tilde{N'}},
\end{eqnarray*}
where $\tilde{N'}\equiv 5(N'-N_T)+3(N-N_T)$ (property (i) in Remark 1),
\begin{widetext}
\begin{eqnarray}
\nonumber
&&p_{\rm incorrect}\\
\nonumber
&\le&\sum_{\nu_T,{\bf s},{\bf b'},k'}p(\nu_T){\rm Tr}\Bigg[P[\otimes_{t\in T'(\nu)}|r_t\rangle]\Bigg[\sum_{\tilde{j'}}\sum_{j''}\alpha_{k'\tilde{j'}}\alpha_{k'j''}^\ast P[|0\rangle^{\otimes 4N}|{\bf s}\rangle|{\bf b'}\rangle]\sigma_{\tilde{j'}}\\
\nonumber
&&\left(\otimes_tP\left[\sum_{{\bf s}_T}\sqrt{p({\bf s}_T)}|{\bf s}_T\rangle HZ(\delta_t)|B_t\rangle|\delta_t\rangle\right]\right)\otimes(I/2)^{\otimes\tilde{N'}}\sigma_{j''}\Bigg]\Bigg]\\
\nonumber
&=&\sum_{\nu_T,k'}p(\nu_T){\rm Tr}\Bigg[P[\otimes_{t\in T'(\nu)}|r_t\rangle]\Bigg[\sum_{\tilde{j'}}\sum_{j''}\alpha_{k'\tilde{j'}}\alpha_{k'j''}^\ast P[|0\rangle^{\otimes 4N_T}]\sigma_{\tilde{j'}}\\
\label{29}
&&\left(\otimes_tP\left[\sum_{{\bf s}_T}\sqrt{p({\bf s}_T)}|{\bf s}_{T}\rangle HZ(\delta_t)|B_t\rangle\right]\right)\otimes(I/2)^{\otimes\tilde{N'}}\sigma_{j''}\Bigg]\Bigg].
\end{eqnarray}
\end{widetext}
Here, ${\bf s}_T\equiv\{s_{5t-4},s_{5t-3},s_{5t-1},s_{5t}\}$, $|B_t\rangle$ is $|B\rangle$ that is a trap qubit, and $|B_t\rangle$ depends on ${\bf s}_T$. We devide $\nu_T$ into $\{{\bf a}^{(t)},{\bf c}^{(t)}\}$ and its complementary  set ${\nu'}_T$. Since
\begin{widetext}
\begin{eqnarray}
\nonumber
&&\sum_{\{{\bf a}^{(t)},{\bf c}^{(t)}\}}p(\{{\bf a}^{(t)},{\bf c}^{(t)}\})\left(\otimes_t P\left[\sum_{{\bf s}_T}\sqrt{p({\bf s}_T)}|{\bf s}_{T}\rangle HZ(\delta_t)|B_t\rangle\right)\right]\\
\nonumber
&=&\otimes_t\cfrac{1}{4}\Bigg(P\left[\sum_{{\bf s}_T}\cfrac{1}{4}|{\bf s}_T\rangle|r_t\rangle\right]+P\left[\sum_{{\bf s}_T}\cfrac{1}{4}|{\bf s}_T\rangle|r_t\oplus s_{5t-1}\oplus s_{5t}\rangle\right]\\
\nonumber
&&+\cfrac{1}{2}\sum_{\theta'}P\left[\sum_{{\bf s}_T}\cfrac{1}{4}|{\bf s}_T\rangle Z(-\theta's_{5t-3})H^{s_{5t-3}}|r_t\oplus s_{5t-4}\rangle\right]\\
\nonumber
&&+\cfrac{1}{2}\sum_{\theta'}P\left[\sum_{{\bf s}_T}\cfrac{1}{4}|{\bf s}_T\rangle Z(-\theta's_{5t-3})H^{s_{5t-3}}|r_t\oplus s_{5t-4}\oplus s_{5t-3}\oplus s_{5t-1}\oplus s_{5t}\rangle\right]\Bigg)\\
\label{E_1}
&\equiv&\mathcal{E}\left(\otimes_tP\left[\sum_{{\bf s}_T}\cfrac{1}{4}|{\bf s}_T\rangle|r_t\rangle\right]\right),
\end{eqnarray}
\end{widetext}
where $\theta'\in\{\pi/2,3\pi/2\}$, Eq.~(\ref{29}) is calculated as follows:
\begin{widetext}
\begin{eqnarray*}
\nonumber
&&p_{\rm incorrect}\\
\nonumber
&\le&\sum_{{\nu'}_T,k'}p({\nu'}_T){\rm Tr}\Bigg[P[\otimes_{t\in T'(\nu)}|r_t\rangle]\Bigg[\sum_{\tilde{j'}}\sum_{j''}\alpha_{k'\tilde{j'}}\alpha_{k'j''}^\ast P[|0\rangle^{\otimes 4N_T}]\sigma_{\tilde{j'}}\\
&&\mathcal{E}\left(\otimes_tP\left[\sum_{{\bf s}_T}\cfrac{1}{4}|{\bf s}_T\rangle|r_t\rangle\right]\right)\otimes(I/2)^{\otimes\tilde{N'}}\sigma_{j''}\Bigg]\Bigg].
\end{eqnarray*}
\end{widetext}
From Eq.~(\ref{E_1}), $\mathcal{E}$ can be treated as TPCP map that is independent of $r_t$. Accordingly, we can treat $\mathcal{E}$ as Bob's deviation, and we define new operator $\Omega'$ that represents Bob's deviation including $\mathcal{E}$ as follows:
\begin{widetext}
\begin{eqnarray*}
\Omega'(\cdot)&\equiv&{\rm Tr}_T\left[\sum_{k',\tilde{j'},j''}\alpha_{k'\tilde{j'}}\alpha_{k'j''}^\ast P[|0\rangle^{4N_T}]\sigma_{\tilde{j'}}\mathcal{E}\left(\otimes_t \left(\sum_{{\bf s}_T}\cfrac{1}{4}|{{\bf s}_T}\rangle\right)\right)\sigma_{j''}\right](\cdot)\\
&\equiv&\sum_{\tilde{k}}{\chi'}_{\tilde{k}}(\cdot){{\chi'}_{\tilde{k}}}^\dag
\end{eqnarray*}
\end{widetext}
such that ${\chi'}_{\tilde{k}}=\sum_{\tilde{j''}}{\alpha'}_{\tilde{k}\tilde{j''}}\sigma_{\tilde{j''}}$, where $\sum_{\tilde{k},\tilde{j''}}|{\alpha'}_{\tilde{k}\tilde{j''}}|^2\le p_n$.
Here, ${\rm Tr}_T$ represents the partial trace over the space spanned by $\{|{\bf s}_T\rangle\}$. The reason why $\sum_{\tilde{k},\tilde{j''}}|{\alpha'}_{\tilde{k}\tilde{j''}}|^2\le p_n$ is the discarding procedure in the virtual gadget. We denote the action of $\sigma_{\tilde{j'}}$ on the $\gamma$th qubit used in the FK protocol by $\sigma_{\tilde{j''}|\gamma}$ $(1\le\gamma\le N)$, 
and define the sets
\begin{eqnarray*}
A_{\tilde{j''}}&\equiv&\{\gamma\ \ {\rm s.t.}\ \ \sigma_{\tilde{j''}|\gamma}=I\}\\
B_{\tilde{j''}}&\equiv&\{\gamma\ \ {\rm s.t.}\ \ \sigma_{\tilde{j''}|\gamma}=X\}\\
C_{\tilde{j''}}&\equiv&\{\gamma\ \ {\rm s.t.}\ \ \sigma_{\tilde{j''}|\gamma}=XZ\}\\
D_{\tilde{j''}}&\equiv&\{\gamma\ \ {\rm s.t.}\ \ \sigma_{\tilde{j''}|\gamma}=Z\},
\end{eqnarray*}
where $|\cdot|$ denotes the number of elements of a set. Note that we can assume that Bob does not perform the deviation on $|\delta_{i'}\rangle$ without loss of generality.
We define the set of $\tilde{j''}$, 
which satisfies $|B_{\tilde{j''}}|+|C_{\tilde{j''}}|\ge d$, as $E_{\tilde{j''}}$. 
Since $I$ and $Z$ do not affect the outcome of the $Z$-basis measurement, and we assume that an error-correcting code that can correct less than $d$ errors is used in the FK protocol,
\begin{widetext}
\begin{eqnarray*}
\nonumber
&&p_{\rm incorrect}\\
&\le&\sum_{{\nu'}_T,\tilde{k}}p({\nu'}_T){\rm Tr}\Bigg[P[\otimes_{t\in T'(\nu)}|r_t\rangle]\Bigg[\sum_{\tilde{j'}\in E_{\tilde{j''}}}\sum_{j'''\in E_{j'''}}{\alpha'}_{\tilde{k}\tilde{j''}}{\alpha'}_{\tilde{k}j'''}^\ast \sigma_{\tilde{j''}}\left(\otimes_tP\left[|r_t\rangle\right]\right)\otimes(I/2)^{\otimes N-N_T}\sigma_{j'''}\Bigg]\Bigg].
\end{eqnarray*}
\end{widetext}
Since if two sigle-qubit Pauli operators $\sigma$ and $\sigma'$ satisfy that $\sigma\neq\sigma'$, $\sum_{r_t}\langle r_t|\sigma|r_t\rangle\langle r_t|\sigma'|r_t\rangle=0$,
\begin{widetext}
\begin{eqnarray*}
\nonumber
&&p_{\rm incorrect}\\
&\le&\sum_{{\nu'}_T,\tilde{k}}p({\nu'}_T){\rm Tr}\Bigg[P[\otimes_{t\in T'(\nu)}|r_t\rangle]\Bigg[\sum_{\tilde{j'}\in E_{\tilde{j''}}}|{\alpha'}_{\tilde{k}\tilde{j''}}|^2 \sigma_{\tilde{j''}}\left(\otimes_tP\left[|r_t\rangle\right]\right)\otimes(I/2)^{\otimes N-N_T}\sigma_{\tilde{j''}}\Bigg]\Bigg]\\
&=&\sum_{{\nu'}_T,\tilde{k}}\sum_{\tilde{j'}\in E_{\tilde{j''}}}|{\alpha'}_{\tilde{k}\tilde{j''}}|^2p({\nu'}_T)\prod_{t\in T'({\nu'}_T)}(\langle r_t|\sigma_{\tilde{j''}|t}|r_t\rangle)^2\\
&=&\sum_{\tilde{k}}\sum_{\tilde{j'}\in E_{\tilde{j''}}}|{\alpha'}_{\tilde{k}\tilde{j''}}|^2\sum_{T'}p(T')\prod_{t\in T'}\sum_{r_t=0}^1p(r_t)(\langle r_t|\sigma_{\tilde{j''}|t}|r_t\rangle)^2.
\end{eqnarray*}
\end{widetext}
We assume that $3N_T=N$, and partition the qubits into $N_T$ sets where each of them contains one trap qubit and two non-trap qubits, respectively. In this time, the position of a trap qubit in each set is chosen uniformly random. We define $|r_{t_{\gamma'}}\rangle$ as a state of a trap qubit that is contained in the $\gamma'$th set. Since this partition gives the information about the location of trap qubits, this partition increase $p_{\rm incorrect}$. Accordingly,
\begin{widetext}
\begin{eqnarray*}
\nonumber
&&p_{\rm incorrect}\\
&\le&\sum_{\tilde{k}}\sum_{\tilde{j'}\in E_{\tilde{j''}}}|{\alpha'}_{\tilde{k}\tilde{j''}}|^2\prod_{\gamma'=1}^{N_T}\sum_{t_{\gamma'}}\sum_{r_{t_{\gamma'}}=0}^1p(t_{\gamma'})p(r_{t_{\gamma'}})(\langle r_{t_{\gamma'}}|\sigma_{\tilde{j''}|t_{\gamma'}}|r_{t_{\gamma'}}\rangle)^2\\
&=&\sum_{\tilde{k}}\sum_{\tilde{j'}\in E_{\tilde{j''}}}|{\alpha'}_{\tilde{k}\tilde{j''}}|^2\prod_{\gamma'=1}^{N_T}\sum_{t_{\gamma'}}\sum_{r_{t_{\gamma'}}=0}^1\cfrac{N_T}{2N}(\langle r_{t_{\gamma'}}|\sigma_{\tilde{j''}|t_{\gamma'}}|r_{t_{\gamma'}}\rangle)^2.
\end{eqnarray*}
\end{widetext}
We define $|A_{\tilde{j''}_{\gamma'}}|$ as the nunmber of elements that satisfies the condition of the set $A_{\tilde{j''}}$ in the $\gamma'$th set. From this definition, 
\begin{eqnarray*}
\sum_{\gamma'=1}^{N_T}|A_{\tilde{j''}_{\gamma'}}|=|A_{\tilde{j''}}|.
\end{eqnarray*}
This definition is applied for other sets $B_{\tilde{j''}}$, $C_{\tilde{j''}}$, and $D_{\tilde{j''}}$. From this definition,
\begin{widetext}
\begin{eqnarray*}
\nonumber
&&p_{\rm incorrect}\\
&\le&\sum_{\tilde{k}}\sum_{\tilde{j'}\in E_{\tilde{j''}}}|{\alpha'}_{\tilde{k}\tilde{j''}}|^2\prod_{\gamma'=1}^{N_T}\cfrac{N_T}{2N}2(|A_{\tilde{j''}_{\gamma'}}|+|D_{\tilde{j''}_{\gamma'}}|)\\
&=&\sum_{\tilde{k}}\sum_{\tilde{j'}\in E_{\tilde{j''}}}|{\alpha'}_{\tilde{k}\tilde{j''}}|^2\prod_{\gamma'=1}^{N_T}\cfrac{N_T}{N}\left(\cfrac{N}{N_T}-|B_{\tilde{j''}_{\gamma'}}|-|C_{\tilde{j''}_{\gamma'}}|\right)\\
&=&\sum_{\tilde{k}}\sum_{\tilde{j'}\in E_{\tilde{j''}}}|{\alpha'}_{\tilde{k}\tilde{j''}}|^2\prod_{\gamma'=1}^{N_T}\left[1-\cfrac{N_T}{N}(|B_{\tilde{j''}_{\gamma'}}|+|C_{\tilde{j''}_{\gamma'}}|)\right].
\end{eqnarray*}
\end{widetext}
From the fact that $(1-gf)\le (1-g)^f$ is satisfied for any non-negative integer $f$ and any real number $g$,
\begin{eqnarray}
\nonumber
&&p_{\rm incorrect}\\
\nonumber
&\le&\sum_{\tilde{k}}\sum_{\tilde{j'}\in E_{\tilde{j''}}}|{\alpha'}_{\tilde{k}\tilde{j''}}|^2\prod_{\gamma'=1}^{N_T}\left(1-\cfrac{N_T}{N}\right)^{|B_{\tilde{j''}_{\gamma'}}|+|C_{\tilde{j''}_{\gamma'}}|}\\
\nonumber
&=&\sum_{\tilde{k}}\sum_{\tilde{j'}\in E_{\tilde{j''}}}|{\alpha'}_{\tilde{k}\tilde{j''}}|^2\left(1-\cfrac{N_T}{N}\right)^{\sum_{\gamma'=1}^{N_T}|B_{\tilde{j''}_{\gamma'}}|+|C_{\tilde{j''}_{\gamma'}}|}\ \ \ \ \ \ \ \\
\nonumber
&=&\sum_{\tilde{k}}\sum_{\tilde{j'}\in E_{\tilde{j''}}}|{\alpha'}_{\tilde{k}\tilde{j''}}|^2\left(1-\cfrac{N_T}{N}\right)^{|B_{\tilde{j''}}|+|C_{\tilde{j''}}|}\\
\nonumber
&\le&\sum_{\tilde{k}}\sum_{\tilde{j'}\in E_{\tilde{j''}}}|{\alpha'}_{\tilde{k}\tilde{j''}}|^2\left(1-\cfrac{N_T}{N}\right)^d\\
\label{ver.}
&\le&p_n\left(1-\cfrac{N_T}{N}\right)^d.
\end{eqnarray}
Since we assume that $3N_T=N$, from Eq.~(\ref{ver.}), 
\begin{eqnarray*}
{p'}_{\rm incorrect}\le\left(\cfrac{2}{3}\right)^d. 
\end{eqnarray*}
\hspace{\fill}$\blacksquare$


\begin{thebibliography}{99}
\bibitem{[FN1]}Since our gadget generates ten kinds of states in Eq.~(\ref{ten}) probabilistically, there is possibility where the number of $Z$-basis states or $\{|+_k\rangle\}$ is more than $N_D$ or $(N-N_D)$, respectively. If so, Alice uniformly randomly selects which extra qubits will be discarded. Then, she instructs Bob to discard them.
\end{thebibliography}
\end{document}